\documentclass[shortbibliography,twocolumn,prl,aps,superscriptaddress,showpacs,amsmath,amssymb,floatfix]{revtex4-2}
\usepackage{graphicx}
\usepackage{amssymb}
\usepackage{amsmath}
\usepackage{epsfig}
\usepackage{color}
\usepackage{mathtools}
\usepackage[colorlinks,linkcolor=blue,anchorcolor=blue,citecolor=blue,urlcolor=blue]{hyperref}
\usepackage{array}
\usepackage{physics}
\usepackage{siunitx}
\usepackage{bm}
\begin{document}

\title{Lorentz Skew Scattering and Giant Nonreciprocal Magneto-Transport}

\author{Cong Xiao}
\email{congxiao@fudan.edu.cn}
\affiliation{Interdisciplinary Center for Theoretical Physics and Information Sciences (ICTPIS), Fudan University, Shanghai 200433, China}

\author{Yue-Xin Huang}
\thanks{These authors contributed equally}
\affiliation{School of Sciences, Great Bay University, Dongguan 523000, China}
\affiliation{Great Bay Institute for Advanced Study, Dongguan 523000, China}

\author{Shengyuan A. Yang}
\affiliation{Research Laboratory for Quantum Materials, IAPME, University of Macau, Taipa, Macau, China}

\begin{abstract}
Skew scattering is the well-known dominant mechanism for anomalous Hall transport in highly conductive systems. However, despite extensive research, the primary mechanism governing nonlinear (nonreciprocal) magneto-transport in clean samples remains unknown. This theoretical gap has impeded the development of design principles for efficient nonreciprocal devices.
Here, we unveil a hitherto unexplored effect in nonreciprocal magneto-transport from cooperative action of Lorentz force and skew scattering. The significance of this Lorentz skew scattering mechanism lies in that it dominates both longitudinal and transverse
responses in highly conductive systems, and it exhibits a scaling behavior distinct from all known mechanisms. 
At low temperature, it shows a cubic scaling in linear conductivity, whereas the scaling becomes quartic at elevated temperature when phonon scattering kicks in. Applying our developed microscopic theory to surface transport in topological crystalline insulator SnTe and bulk transport in Weyl semimetals leads to significant results, suggesting a new route to achieve giant transport nonreciprocity in high-mobility materials with topological band features.

\end{abstract}

\maketitle

Nonreciprocal transport phenomena have received significant attention, as they manifest intriguing physics of electronic quantum geometry and form the basis for rectification and diode applications~\cite{Rikken2001,Tokura2018nonreciprocal,Ideue2021Review}. Particularly, in nonmagnetic crystals with
broken inversion symmetry, an applied magnetic field could trigger a nonreciprocal magneto-resistance linear in the $B$ field \cite{pop2014MchA,iwasa2017,Zhang2018bilinear,he_bilinear_2018,He2018STO,Rikken2019Te,Choe2019gate,Fert2020Ge,Fert2020theory,Zhang2021GeTe,Hueso2022Te,Ando2022nanowire,Wu2022Sn,Wang2022bilinear,Dagan2024,he2019,He2019WTe2,Wang2021NPHE,Kasai2021,Dimi2023hole,huang_Intrinsic_2023,Dantas2023,Ye2023Te}. The corresponding nonreciprocal magneto-transport (NRMT) response current can be expressed
as $j=\chi E^2 B$, with $\chi$ denoting the response tensor. This phenomenon was first studied in chiral crystals (known as electrical magnetochiral anisotropy) \cite{Rikken2001} and recently actively explored also in various achiral crystals \cite{Ando2022ZrTe5,Ando2023ZrTe5,Wang2024ZrTe5,Zhang2024light,Song2024observation}.

In experiment, to understand microscopic mechanisms of a transport phenomenon, a common practice is to perform a scaling analysis, i.e., to analyze how the response coefficient varies as a function of the linear conductivity $\sigma_{xx}$, which is proportional to the scattering time $\tau$. Till now, several mechanisms for NRMT were proposed. For example, an intrinsic mechanism independent of $\sigma_{xx}$ ($\tau$) was revealed for NRMT Hall response \cite{huang_Intrinsic_2023,Huang2023PRB}. For longitudinal response, $\chi$ may originate from chiral scatterer~\cite{Rikken2001}, magnetic self-field~\cite{Rikken2001}, Zeeman-coupling induced Fermi surface deformation \cite{iwasa2017,Okumura2024}, energy relaxation \cite{Spivak2023MchA}, chiral anomaly \cite{Morimoto2016} and Berry curvature \cite{Yokouchi2023WTe2} mechanisms in Weyl semimetals, and etc \cite{Amit2022EEB2D,Wang2023interband,Ma2024nonreciporcal}.
It is noted that: (1) theoretical formulations of the various mechanisms are so far limited within the simple relaxation time approximation which does not fully capture the quantum nature of scattering, and (2) they (except the intrinsic one) all give a  $\chi\propto \sigma_{xx}^2$ scaling.
Thereby, a natural question arises: Is there any mechanism for NRMT, with distinct scaling behavior, from quantum effects
in scattering? Furthermore, one may ask: What mechanism gives the \emph{highest power} in the scaling relation? This is important, because such contribution is expected to dominate the response in clean samples with large $\tau$.

The above mentioned fundamental gap in our understanding of NRMT has hindered the discovery of design principles for efficient, low-power nonreciprocal devices. While skew scattering dominates $B$-field-free nonlinear transport in highly conductive materials like graphene superlattices~\cite{Fu2020}, leading to strong frequency doubling and energy harvesting~\cite{He2022graphene,Duan2022PRL}, the primary NRMT mechanism in such systems remains elusive. Resolving this could not only reveal a hidden field-induced nonreciprocal transport mechanism but also unlock novel pathways to significant nonreciprocity and rectification capabilities.

\begin{figure}
    \centering
    \includegraphics[width=0.48\textwidth]{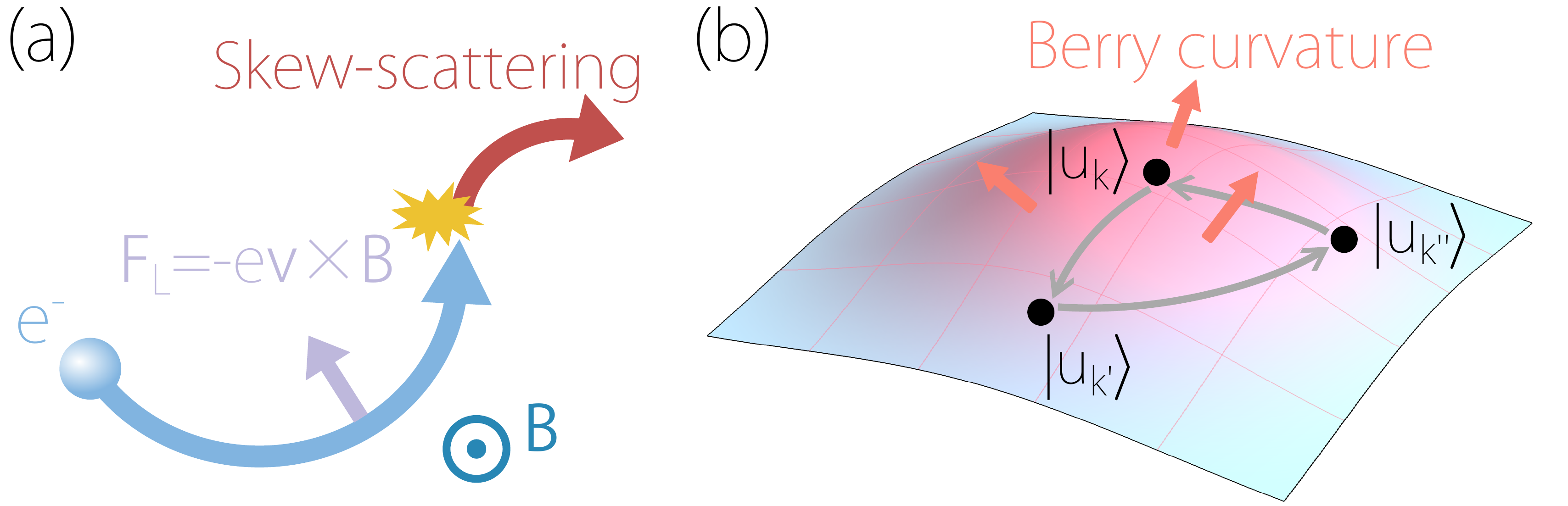}
    \caption{(a) Schematic of actions of Lorentz force and skew scattering on electron motion. 
    (b) Quantum geometric character of skew scattering process, as
    exemplified by a Wilson loop involving three states on Fermi surface. The corresponding skew scattering rate is proportional to the Berry curvature flux through the loop.
    }
    \label{fig-process}
\end{figure}

In this work, we resolve the above issues by unveiling a new mechanism for NRMT --- Lorentz skew scattering (LSK),  which is resulted from the cooperative action of skew scattering (a quantum effect of scattering which induces trajectory skewness) and Lorentz force by magnetic field, as sketched in Fig.~\ref{fig-process}. This mechanism does not require spin-orbit coupling, and it manifests Berry curvature on Fermi surface. Importantly, we show that LSK is the leading contribution with the highest degree in the scaling relation for good metals. Specifically, at low temperatures when impurity scattering dominates, it gives $\chi\propto \sigma_{xx}^3$ scaling; whereas at elevated temperatures when phonon scattering becomes substantial, the LSK contribution would scale as $\sigma_{xx}^4$. Because of its distinct scaling and quantum geometric character, it should be dominating in highly conductive samples and strongly enhanced by topological band features around Fermi level.
We demonstrate our theory by studying surface transport in topological crystalline insulator (TCI) SnTe and bulk transport in Weyl semimetals. The estimated LSK contribution can be orders of magnitude larger than previously studied mechanisms.
As the NRMT in most reported works is rather weak, our finding offers a new insight for amplifying this nonreciprocal effect, which is promising for low-dissipative rectification applications.

\textit{\color{blue} Geometric and scaling characters of LSK.} We consider a diffusive system under weak applied $E$ and $B$ fields in the semiclassical regime. The electric current is generally expressed as
$
    \bm j=-e\sum_l f_l \bm v_l,
$
where $-e$ is the electron charge, $l=(n,\bm k)$ is a collective index labeling a Bloch state,
$f$ is the distribution function, and $\bm v$ is the electron velocity.
To study NRMT response, we focus only on the part of the current $\propto E^2B$.
For our proposed LSK mechanism, $B$ field enters via Lorentz force, while skew scattering enters via the collision integral. They together generate an out-of-equilibrium distribution $f^\text{LSK}\propto E^2B$. (Hence, in calculating the current, it is sufficient to take $v_l$ as the unperturbed band velocity.) This $f^\text{LSK}$ can be obtained from the
Boltzmann kinetic equation:
%
%
%
\begin{align}
(\mathcal{\Hat{D}}_{\text{E}}+\mathcal{\Hat{D}}_{\text{L}}) f_l= (\mathcal{\Hat{I}}_{\text{c}}+\mathcal{\Hat{I}}_{\text{sk}}) f_l
    \label{eq-DD},
\end{align}
where hat denotes linear operators, $\mathcal{\Hat{D}}_{\text{E}}=-\frac{e}{\hbar}\bm E \cdot \partial_{\bm k}$ and $\mathcal{\Hat{D}}_{\text{L}}=-\frac{e}{\hbar} (\bm v_l \cross \bm B) \cdot \partial_{\bm k}$ give the electric force and Lorentz force driving terms, $\mathcal{\Hat{I}}_{\text{c}}$ and $\mathcal{\Hat{I}}_{\text{sk}}$ correspond to the conventional collision integral
and the skew-scattering collision integral, respectively \cite{sinitsyn_semiclassical_2007} (see Methods). For scaling analysis, one can take
$\mathcal{\Hat{I}}_{\text{c}}\sim 1/\tau$ and $\mathcal{\Hat{I}}_{\text{sk}}\sim 1/\tau_\text{sk}$, where
the skew scattering time $\tau_\text{sk}$ should be much larger than $\tau$ \cite{Sinitsyn2006graphene,he_quantum_2021,Fu2020,He2022graphene}.

The leading contribution to skew scattering is from third-order on-shell scattering processes. Assuming weak spin-independent disorders, the skew scattering rate $\omega_{l'l}^{a}$
is related to the Wilson loop connecting the three involved electronic states $l$, $l'$, and $l''$ (schematics in Fig.~\ref{fig-process}(b)):
$
W(l,l',l'')=\bra{u_l}\ket{u_{l'}}\bra{u_{l'}}\ket{u_{l''}}\bra{u_{l''}}\ket{u_{l}}.
$
This quantity is
associated with the Pancharatnam-Berry phase $\text{arg}(W)$ \cite{Vanderbilt_2018}.
For an infinitesimal Wilson loop in $k$ space, one finds
\begin{equation}
  \operatorname{Im}W(l,l',l'')\approx \frac{1}{2}(\bm{k}''-\bm{k}')\times (\bm{k}'-\bm{k})\cdot \bm\Omega_l,
\end{equation}
which is proportional to the Berry curvature $\bm\Omega$. It follows that for an isotropic model with smooth disorder potential,
$\omega_{l'l}^{a}\propto \bm{k}'\cdot (\bm{k}\times \bm \Omega_l)$, explicitly showing the skewness of scattering, i.e.,
an incident electron with momentum $\bm k$ tends to be scattered to the transverse direction
$\bm k\times \bm \Omega_l$. Since the Lorentz force also deflects electrons, their cooperative action
should affects both longitudinal and transverse current flows.
And since for transport, scattering events occur mainly around Fermi level, one expects that skew scattering and hence LSK mechanism would be enhanced if there is substantial Berry curvature distribution on Fermi surface.

Before detailed analysis, one may argue the scaling behavior of $f^\text{LSK}$ in an intuitive way.
The $\tau$ dependence of out-of-equilibrium distribution is associated with the driving field. $E$ field conventionally gives a $\propto\tau$ dependence, but with skew scattering, it leads to an additional contribution $\propto \tau^2/\tau_\text{sk}$, which has a higher degree in $\tau$ and is well known in the study of anomalous Hall effect \cite{nagaosa2010}. For $f\propto E^2B$, if each factor of $E$ is associated with conventional scattering and gives a $\tau$ factor, the resulting distribution would be $\propto \tau^2$. This just corresponds to the previously studied mechanisms \cite{iwasa2017,Zhang2018bilinear,he_bilinear_2018,He2019WTe2,Dagan2024}. Note that in those cases, $B$ field
only enters via correction of band structure and cannot bring additional $\tau$ factors, because by itself a $B$ field cannot drive a nonequilibrium.
This also applies if one lets each $E$ associated with skew scattering, then the result would scale as $(\tau^2/\tau_\text{sk})^2$. In comparison, a much larger contribution arises if $B$ enters via Lorentz force: Combined with a factor of $E$, they together bring a $\tau^2$ factor, which just corresponds to the ordinary Hall conductivity $\propto \tau^2$ \cite{Ziman}. Then, combined with one skew scattering for the remaining $E$ factor, we have a result $\propto \tau^4/\tau_\text{sk}$, which is the LSK contribution we are looking for. This analysis clarifies why LSK can have a higher $\tau$ ($\sigma_{xx}$) scaling than other mechanisms, which implies that LSK should dominate the NRMT in highly conductive materials (small impurity concentration).

\begin{figure}[t]
    \centering
    \includegraphics[width=0.43\textwidth]{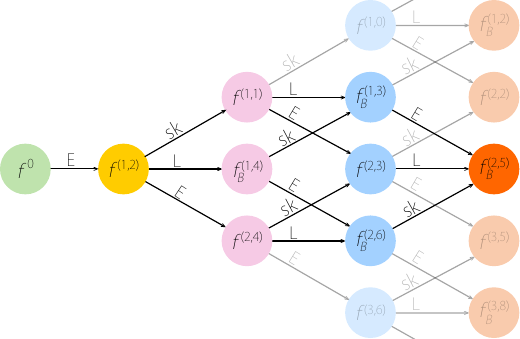}
    \caption{Diagrammatic approach to solve the kinetic equation. Each node denotes a component $f^{(i,j)}$, and each arrow denotes an operation (rules are described in the text). Nodes in each column share the same dependence on $\tau$. 
    From left to right, the components are $\propto \tau^0$, $\tau^1$, $\tau^2$, $\tau^3$ and $\tau^4$, respectively.
    The components relevant to LSK are highlighted. 
    }
    \label{fig-distributionF}
\end{figure}

\textit{\color{blue} Diagrammatic approach to LSK.}
To derive the formula for LSK contribution from Eq.~(\ref{eq-DD}), we use the method of successive approximations \cite{sinitsyn_semiclassical_2007,Xiao2019NLHE,huang_scaling_2025} (Supplementary Note 1) and expand the distribution function as
\begin{equation}\label{ff}
  f=f^0+\sum_{i,j}\Big [f^{(i,j)}
  +f_B^{(i,j)}\Big ].
\end{equation}
Here, $f^0$ is the equilibrium Fermi-Dirac distribution. In the off-equilibrium part, we explicitly separate out the
components $f_B$ which are linear in $B$, and
to keep track of the degrees in $E$ field and scattering potential $V$, we use the notation $Q^{(i,j)}$ to indicate a quantity  $\propto E^i V^{-j}$.
Figure \ref{fig-distributionF} illustrates the diagrammatical way to construct each $f^{(i,j)}$. The \emph{rules} are: (1) Each node is a component of distribution function, and the construction starts from $f^0$; (2) An arrow with label $O$ pointing from node $A$ to $B$ means $f^B$ has a contribution from $f^A$ acted by the operator $\hat O$, and the degrees of $E$, $B$, $V$ must be balanced between $f^B$ and $\hat O f^A$; (3) Here, we have three types of arrow labels with the correspondence: $E\rightarrow -\tau\mathcal{\Hat{D}}_{\text{E}}$,  $L\rightarrow -\tau\mathcal{\Hat{D}}_{\text{L}}$, and $sk\rightarrow \tau \mathcal{\Hat{I}}_{\text{sk}}$. (4) The component at a node is obtained by summing all contributions associated with arrows pointing to it. In addition,
there is no arrow from $f^0$ with $L$ or $sk$ label, since they cannot produce nonequilibrium distribution without $E$. Following these rules, one can readily obtain any desired component $f^{(i,j)}$.

Our target is to solve $f^\text{LSK}$, which is at the second order of electric field and involves one Lorentz force action ($\mathcal{\Hat{D}}_{\text{E}}\mathcal{\Hat{D}}_{\text{L}}$) and one skew scattering
($\mathcal{\Hat{D}}_{\text{E}}\mathcal{\Hat{I}}_\text{sk}$).
According to the diagrammatic approach, we identify it as $f_B^{(2,5)}$ in
Fig.~\ref{fig-distributionF}. Its expression can thus be read off from the diagram as
\begin{widetext}
\begin{align}
     f^\text{LSK}=f_B^{(2,5)}= -\tau^4 \Big [\mathcal{\Hat{D}}_{\text{E}}\{\mathcal{\Hat{D}}_{\text{L}},\mathcal{\Hat{I}}_{\text{sk}}\}
    +\mathcal{\Hat{D}}_{\text{L}}\{\mathcal{\Hat{I}}_{\text{sk}},\mathcal{\Hat{D}}_{\text{E}}\}
    +\mathcal{\Hat{I}}_{\text{sk}}\{\mathcal{\Hat{D}}_{\text{E}},\mathcal{\Hat{D}}_{\text{L}}\}
    \Big ] \mathcal{\Hat{D}}_{\text{E}} f^0.
    \label{eq-g25All}
\end{align}
\end{widetext}
Here, $\acomm{.}{.}$ is the anticommutator of two operators. Combined with the band velocity, it gives the LSK response current in NRMT:
$\bm j^{\text{LSK}}=-e \sum_l f^\text{LSK}_l \bm v_l$,
from which the response tensor $\chi^\text{LSK}$ can be extracted ($j^{\mathrm{LSK}}_{a}=\chi^{\text{LSK}}_{abcd}E_b E_cB_d$, where summation over repeated Cartesian indices is implied).

From Eq.~(\ref{eq-g25All}), the scaling behavior $f^\text{LSK}, \chi^\text{LSK}\propto\tau^4/\tau_\text{sk}$ is consistent with our previous analysis. However, to discuss the scaling of
$\chi^\text{LSK}$ with $\sigma_{xx}$, we have to distinguish two regimes. In the low temperature regime where impurity scattering dominates, $\sigma_{xx}$ is usually varied by fabricating samples with varying impurity density $n_i$. For example, in Refs. \cite{hou2015,Jin2017}, this is done by making metal films with different thicknesses such that the effective $n_i$ from surface scattering is varied. Since both $\tau$ and $\tau_\text{sk}$ are $\propto 1/n_i$, we expect for such cases, $\chi^\text{LSK}\propto \sigma_{xx}^3$. On the other hand, at elevated temperatures where phonon scattering is substantial,
$\tau$ (and $\sigma_{xx}$) is usually varied by temperature, due to phonon scattering. Meanwhile, phonons do not contribute to skew scattering \cite{hou2015,lyo_ferromagnetic_1973,yang_scattering_2011}, so $\tau_\text{sk}$ is still from impurity scattering. Therefore, one should observe $\chi^\text{LSK}\propto \sigma_{xx}^4$. These scaling behaviors are distinct from all previous mechanisms for NRMT.



\begin{figure}
    \centering
    \includegraphics[width=0.45\textwidth]{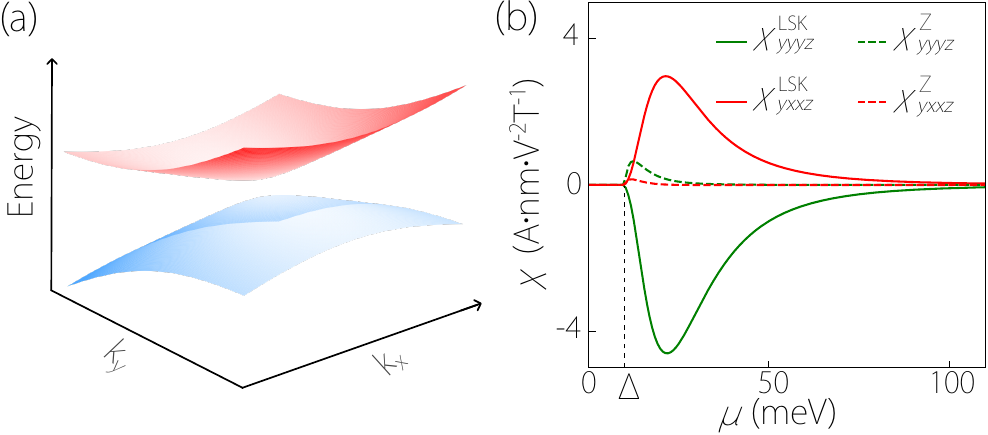}
    \caption{(a) Dispersion of a 2D gapped Dirac valley in (\ref{2D model}). (b) Calculated LSK nonlinear conductivities versus chemical potential for this model. For comparison, the dashed lines ($\chi^{Z}$'s) show the contribution from mechanism of Fermi surface deformation by Zeeman coupling to orbital moment. Here, we take parameters relevant to SnTe, with $v_x/\hbar=v_y/\hbar=\SI{4e5}{m/s}$, $\Delta=\SI{10}{meV}$, $w/v=0.1$, $n_i=\SI{e10}{cm^{-2}}$, and averaged disorder strength $V_0=\SI{e-13}{eV\cdot cm^2}$.}
    \label{fig-2Dmodel}
\end{figure}

\textit{\color{blue}Giant LSK response in surface states of TCI.}
We first apply our theory to the 2D Dirac model:
\begin{align}
    H= \tau wk_y+v_xk_x \sigma_y-\tau v_y k_y \sigma_x+\Delta \sigma_z,
    \label{2D model}
\end{align}
where $\tau=\pm$ labels two Dirac valleys connected by time reversal operation $\mathcal{T}$, and $\sigma$'s are Pauli matrices. This model describes the surface states of TCIs SnTe \cite{Ando2012SnTe} and Pb$_{1-x}$Sn$_x$Te(Se) \cite{okada_observation_2013} at low temperatures. The spectrum for one valley is shown in Fig.~\ref{fig-2Dmodel}(a).

\renewcommand{\arraystretch}{1.5}
\begin{table*}
\caption{Comparison of reported nonreciprocal coefficients $\gamma$ and $\gamma'$ with different mechanisms and materials. For 3D systems, the intrinsic coefficient $\gamma'$ independent of sample size is also presented for comparison.}
    \centering
    \begin{tabular}{p{0.11\linewidth}>{\centering\arraybackslash}p{0.25\linewidth}<{\centering}p{0.2\linewidth}<{\centering}p{0.20\linewidth}<{\centering}p{0.20\linewidth}<{\centering}}
        \hline\hline
        Ref. & Mechanism & Platform & $\gamma\,(\si{T^{-1}A^{-1}})$ & $\gamma'\,(\si{m^2T^{-1}A^{-1}})$
        \\
        \hline
        \cite{Rikken2001} & Chiral scatterer, magnetic self-field & Bi helix & $10^{-3}$ & $10^{-10}$
        \\
        \cite{pop2014MchA} &  Not identified & $\mathrm{[DM-EDT-TTF]_2ClO_4}$ & $10^{-2}$ & $10^{-10}$
         \\
        \cite{Morimoto2016} & Chiral anomaly & TaAs & $0.3$ & $3\times 10^{-8}$
        \\
        \cite{iwasa2017} &  Zeeman-coupling induced Fermi surface deformation  & $\mathrm{BiTeBr}$ & $-$ & $3\times 10^{-12}$
        \\
        \cite{he_bilinear_2018} &  Zeeman-coupling induced Fermi surface deformation  & $\mathrm{Bi_2Se_3}$ & $\approx 2\times 10^{-3}$ & $\approx 4\times 10^{-17}$
        \\
        \cite{He2018STO} &  Not identified & $\mathrm{SrTiO_3(111)}$ surface & $20$ & $-$
         \\
        \cite{Choe2019gate} & Not identified & $\mathrm{LaAlO_3/SrTiO_3}$ interface & $10^2$ & $-$
        \\
        \cite{Ando2022ZrTe5} & Not identified & $\mathrm{ZrTe_5}$ & $-$ & $4\times 10^{-7}$
        \\
        \cite{Ando2022nanowire} & special dispersion of quantum-confined surface state & $\mathrm{(Bi_{1-x}Sb_x)_2 Te_3}$ nanowire & $10^5$ & $-$
        \\
        \cite{Yokouchi2023WTe2} & Berry curvature & $\mathrm{WTe_2}$ & $-$ & $3.4\times 10^{-7}$
        \\ 
        $\bm {\mathrm{This \ work}}$ & $\bm {\mathrm{LSK}}$ & surface of SnTe & $ 4\times 10^{5}$ & $-$
        \\ 
        $\bm {\mathrm{This \ work}}$ & $\bm {\mathrm{LSK}}$ & clean Weyl semimetal & $-$ & $4\times 10^{-6}$
        \\ \hline\hline
    \end{tabular}
    \label{tab-comparison}
\end{table*}

To have Lorentz force, we take $B$ field to be in the $z$ direction. By Eq.~\eqref{eq-g25All}, near the bottom of the upper Dirac band, we estimate that the $\chi^\text{LSK}$ components for both longitudinal and transverse NRMT can reach a similar order of magnitude, with (Supplementary Note 2)
\begin{equation}
  \left|\chi^\text{LSK}\right|\sim \frac{e^4 w}{\pi^2\hbar^5 D} \frac{\tau^4}{\tau_\text{sk}},
\end{equation}
where $D$ is the density of states. The results from numerical calculations (using parameters of SnTe \cite{Ando2012SnTe,sodemann_quantum_2015}) are plotted in Fig.~\ref{fig-2Dmodel}(b), which exhibit a peak near band bottom, because of the sizable Berry curvature in this region. In the figure, for comparison, we also plot the results from the mechanism of Fermi surface deformation by Zeeman coupling to orbital moment ($\chi^Z$) \cite{Okumura2024}, which are found to be much smaller than the LSK mechanism.

The nonreciprocity is often characterized by the coefficient
$
    \eta =\delta\sigma/\sigma= -\delta \rho /\rho,
$
measuring the change in conductivity (resistivity) when the current direction is reversed. Here, we find $\eta$ from LSK can reach $\sim 20\%$ under $B=1$ T and $E=10^4$ V/m. This value is orders of magnitude larger than several previous results of
NRMT in 2D electron gas under similar field strengths \cite{he_bilinear_2018,He2018STO,Choe2019gate}. Another figure of merit is the nonreciprocal coefficient $\gamma=-\eta/IB$, where $I$ is the driving current \cite{Tokura2018nonreciprocal}. For a sample width of 1 $\mu$m, we estimate 
$\gamma$ here can reach $\sim 10^5$ A$^{-1}$T$^{-1}$, which is very large, considering that most reported $\gamma$ values are below $10^3$ A$^{-1}$T$^{-1}$ \cite{Rikken2001,pop2014MchA,iwasa2017,he_bilinear_2018,He2018STO,Rikken2019Te,Choe2019gate}.

\textit{\color{blue} Giant LSK nonreciprocity in Weyl semimetal.} We have shown that to have pronounced LSK response, the system should have high mobility and large Berry curvature on Fermi surface. Weyl semimetals satisfy these conditions \cite{Armitage2018}. In a Weyl semimetal, the low-energy physics is from states around Weyl points~\cite{Armitage2018}. A generic model for a Weyl point can be written as
\begin{align}
H=wk_z+v\bm k\cdot \bm{\sigma},
\label{eq-3DH}
\end{align}
which acts as a monopole for Berry curvature field. Since LSK contribution is $\mathcal{T}$-even, a pair of Weyl points connected by $\mathcal{T}$ give the same contribution.

\begin{figure}[t]
    \centering
    \includegraphics[width=0.48\textwidth]{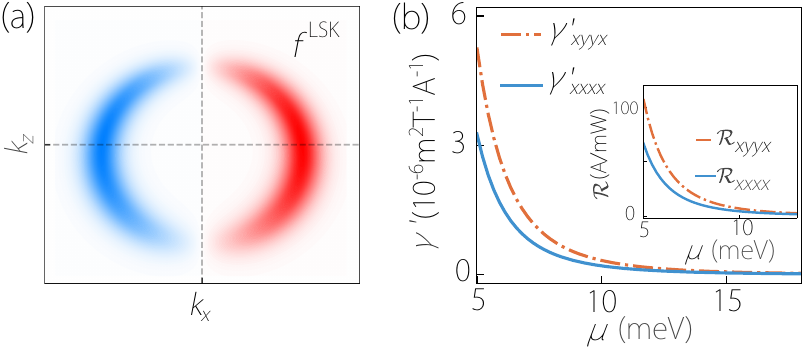}
    \caption{LSK response for the Weyl model in Eq.~(\ref{eq-3DH}). (a) Patterns of $f^\text{LSK}$ on the Fermi surface in $k_y=0$ plane, for $E$ and $B$ fields applied in $x$ direction and $\mu=\SI{10}{meV}$. (b) Calculated nonreciprocal coefficient $\gamma'$ versus chemical potential. The inset shows the obtained current responsivity. In the calculation, we take $B=\SI{0.1}{T}$, $v=\SI{4e5}{m/s}$, $w/v=0.4$,
    $n_i=\SI{e15}{cm^{-3}}$, and $V_0=\SI{e-19}{eV\cdot cm^3}$.}
    \label{fig-model3D}
\end{figure}

We perform numerical calculation for this Weyl model using parameters typical of Weyl semimetal materials (such as TaP family~\cite{grassano_influence_2020}). Figure~\ref{fig-model3D}(a) illustrates the obtained $f^\text{LSK}$ distribution. 
For bulk materials, one usually characterizes NRMT using an intrinsic coefficient $\gamma'=\gamma A=-\chi/\sigma_{xx}^2$, where $A$ is the cross sectional area of the sample \cite{Morimoto2016,iwasa2017,Ando2022ZrTe5}. Figure~\ref{fig-model3D}(b) shows the result.
One finds that $\gamma'$ can reach $3\times 10^{-6}$ m$^2$A$^{-1}$T$^{-1}$ for $\mu=5$ meV above Weyl point. Such LSK contribution is at least an order of magnitude larger than the chiral anomaly contribution and other mechanisms previously proposed \cite{Morimoto2016}. This demonstrates LSK could dominate the NRMT response in Weyl semimetals.



\textit{\color{blue}Discussion.} The proposed LSK mechanism for NRMT is significant because it manifests quantum geometry of band structure (Berry curvature on Fermi surface) and is dominant in highly conductive samples (possessing the highest scaling power in the Drude conductivity). The comparisons of LSK and previously reported NRMT are presented in Table.~\ref{tab-comparison}. 
One sees that LSK can surpass the other known mechanisms by orders of magnitude.
As we noted,
materials with topological band features around Fermi level, such as topological semimetals, should be ideal systems to study LSK. Recently, signals of strong skew scattering effects in nonlinear Hall measurement were reported in
several systems, such as graphene superlattices~\cite{He2022graphene,Duan2022PRL}, BiTeBr~\cite{Lu2024BiTeBr} and Te thin flakes~\cite{Zeng2024giant}. They could be promising platforms to explore LSK response as well.

The revealed strength of LSK transport also highlights the previously overlooked significance of the classical Lorentz-force effect of magnetic field on NRMT. In highly conductive materials, the LSK from classical-quantum mixture (Lorentz force and skew scattering) should naturally dominate over purely quantum effects. To see this, we compare LSK with the strongest quantum nonlinear effect $\propto E^2B$. The latter is given by Zeeman corrected second order skew scattering, i.e., composition of two skew scattering processes~\cite{huang_scaling_2025} (Supplementary Note 3), which gives a NRMT contribution scaling as $\sigma_{xx}^2$ assuming static impurity scattering thus should be small compared to LSK in highly conductive samples. We also show this comparison explicitly by using model (\ref{2D model}): $\chi$ and $\gamma$ due to the $B$-corrected second order skew scattering are two orders of magnitude smaller than LSK. 

The diagrammatic approach developed in this study offers a general method to tackle the Boltzmann equation for
nonlinear transport. The various processes involved in a response can be easily identified and intuitively visualized.
Via this approach, we also find contributions from higher-order LSK processes, which are much smaller than $f^\text{LSK}$ by factors $(\tau/\tau_\text{sk})^2$ and $(\tau/\tau_\text{sk})^4$, so they can generally be neglected.

The LSK mechanism is not limited to electrical transport but should also affect other processes, such as nonreciprocal thermal and thermoelectric transport. In particular, it may play a significant role in thermal rectification \cite{Walker2011review,Li2012RMP}, which is an important direction of research.

The LSK induced NRMT is suitable for rectifier or detector applications, since such devices require high mobility materials, which could reduce power consumption and heat dissipation. An important metric for rectification applications is the current responsivity $\mathcal{R}=j_{dc}/P$, which is the ratio of the output dc current to the power dissipation $P$ \cite{zhang2021}.
For the Weyl semimetal case, we estimate that $\mathcal{R}$ due to LSK may reach $\sim 66$ A/mW at $\mu=5$ meV, for $B=0.1$ T and a device size of $1~\mu$m, as shown in the inset of Fig.~\ref{fig-model3D}(b). This value is already orders of magnitude larger than other reported rectification mechanisms~\cite{zhang2021,Rectification2019Review}. All these suggest that rectification based on LSK indeed holds potential for practical applications.

\section{METHODS}

\subsection{Formulation of Boltzmann equation}

The standard Boltzmann kinetic equation at steady states reads:
\begin{align}
    \dot{\bm k}\cdot \pdv{f_l}{\bm k}=\mathcal{\Hat{I}}\left\{ f_l \right\}.
    \label{eq-BolEq}
\end{align}
According to the semiclassical equation of motion
for a Bloch electron wave packet, $\dot{\bm k}$ can be denoted as two parts $\hat{\mathcal D}_E$ and $\hat{\mathcal D}_L$, as a result
of the electric force and Lorentz force. Because we are considering the LSK process, the magnetic field enters via the Lorentz force, and other $B$-related terms are not relevant. For the collision integral, we do not consider the side jump part and the Zeeman coupling induced $B$-field correction to collision integrals \cite{xiao2020mr}. These effects may contribute to NRMT, but they do not contribute to LSK response and their scaling degree is lower than $\chi^\text{LSK}$.
Therefore, the relevant collsion integral to LSK is $\mathcal{\Hat{I}}=\mathcal I_c+\mathcal{\Hat{I}}_\text{sk}$, where
$\mathcal{\Hat{I}}_{\text{c}}$ and $\mathcal{\Hat{I}}_{\text{sk}}$ correspond to the conventional collision integral
$
  \mathcal{\Hat{I}}_{\text{c}} f_l=-\sum_{l'}\omega_{l'l}^{s}(f_l-f_{l'})
$ and the skew-scattering collision integral
$
  \mathcal{\Hat{I}}_{\text{sk}} f_l=-\sum_{l'}\omega_{l'l}^{a}(f_l+f_{l'})
$, respectively \cite{sinitsyn_semiclassical_2007}. Here, $\omega_{l'l}^{s}$ is the symmetric scattering rate between $l$ and $l'$ states, and $\omega_{l'l}^{a}$ is the antisymmetric (skew) scattering rate.
Gathering the above considerations, Eq.~\eqref{eq-BolEq} reduces to Eq.~\eqref{eq-DD} for investigating the LSK effect.

Assuming weak spin-independent disorders, the skew scattering rate takes the form of
\begin{equation}
    \omega_{l'l}^{a}\approx\frac{4\pi^2}{\hbar}n_i \sum_{l''}\langle V_{ll''}V_{l''l'}V_{l'l}\rangle_c \delta (\varepsilon_{l'l})\delta (\varepsilon_{l''l})\operatorname{Im}{W(l,l',l'')}\nonumber,
\end{equation}
where $n_i$ is disorder density, $V_{ll'}=V_{\bm k-\bm k'}$ is the Fourier component of disorder potential, $\langle ... \rangle_c$ denotes the disorder average, $\varepsilon_{l'l}\equiv \varepsilon_{l'}-\varepsilon_l$,
and $W(l,l',l'')$ is the Wilson loop.

\bibliography{ref}




\end{document}